\documentclass[11pt,twoside]{article}

\usepackage{asp2004hack}
\usepackage{epsf}
\usepackage{lscape}
\usepackage{graphicx}
\usepackage{amssymb}
\begin{document}
\title{A New Look at the Empirical Initial-Final Mass Relation}   
\author{Kurtis A. Williams}   
\affil{Department of Astronomy, University of Texas, 1 University
  Sta., C1400, Austin, TX, 78712}    

\begin{abstract} 
We examine new methods of producing and analyzing the empirical
initial-final mass relation for open cluster white dwarfs (WDs). We
re-determine initial and final masses for the complete sample of
published cluster WDs and then pare this sample using stringent
criteria.  We create an empirical initial-final mass relation by
binning all WDs in individual clusters to a single point.  Despite
potentially significant systematics arising from this approach, we
are comfortable concluding that, to within current observational
constraints, the initial-final mass relation is linear, any intrinsic
scatter in the relation is $\lesssim 0.05M_\odot$, and there is no
metallicity dependence.  More exploration of these issues is clearly
warranted. 
\end{abstract}

\section{Introduction}
The initial-final mass relation (IFMR) represents the integrated mass
lost by a star over its entire evolution from zero-age main sequence
to the white dwarf (WD) cooling sequence.  As such, the IFMR can
provide valuable insight into difficult issues of stellar mass loss.  
The IFMR is an integral part of widely-varied areas of
astrophysical research, from dating the age of the Galactic disk via
the WD luminosity function \citep[e.g.,][]{Winget1987}, to
understanding chemical enrichment and star formation efficiencies in
galaxies \citep[e.g.,][]{Ferrario2005}, to the origin and evolution of hot gas
in elliptical galaxies \citep[e.g.,][]{Mathews1990}.  

The first comparison of measurements of WD masses and
their progenitor masses to theoretical predictions of the IFMR was
made by \citet{Weidemann1977}.  With increasing numbers of open
cluster WD studies and modern instrumentation, the number of published
open cluster WDs has grown to $\sim 50$ \citep{Ferrario2005} and is
rapidly increasing \citep[e.g.,][and J.~Kalirai, this
  volume]{Williams2006}. 

Despite the rapid growth in the number of observational points on the
IFMR, many uncertainties remain regarding the relation.  Is the
relation linear or more complicated?  What is the intrinsic scatter;
i.e., what is the range of WD masses a main-sequence star of a given
mass will produce?  Is there any dependence of the IFMR on
metallicity?  A glance at the current empirical IFMR (see Figure 1)
leads one to understand why answers to these questions have not been
forthcoming -- the scatter in the points is gigantic, masking
any subtle trends.

\begin{figure}[!tb]
\begin{center}
\includegraphics[scale=0.5]{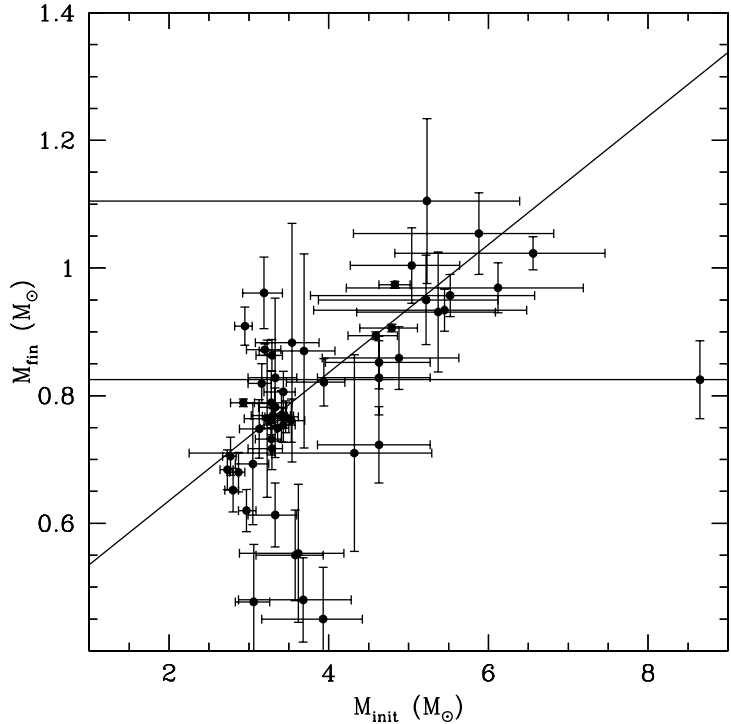}
\end{center}
\caption{The empirical initial-final mass relation from Ferrario et
  al.~(2005), including subsequent Praesepe points from work by
  \citet{Dobbie2006a}.  Error bars include observational errors and
  uncertainties in cluster ages.  The line is the best fit model from
  Ferrario et al.  References for individual points are given in
  Ferrario et al.~(2005). \label{fig.lilia}}
\end{figure}

In order to better understand the IFMR, including its intrinsic
scatter and any metallicity dependence, a more sophisticated means of
of creating and analyzing the empirical IFMR is needed.  We are
beginning an exploration into various means of analyzing the IFMR, and
present one early attempt in these proceedings.  Although the
quantitative conclusions presented herein should be regarded as
preliminary, we can conclude that the current WD sample is of
sufficient quantity and quality to address the issues raised above.
We therefore continue to explore more robust methods of
analyzing the empirical IFMR.

\section{Toward a Robust Empirical Initial-Final Mass Relation}
Given that the current open cluster data come from a wide variety of
data sets with widely varying quality and constraints, our first order
of business is to put all the data on similar footing.  Using
published effective temperatures, surface gravities, and errors, we
re-determine each open cluster WD's mass and cooling age using
evolutionary models graciously provided by P.~Bergeron \citep[For
  description of the models, see][]{Holberg2006}.  Using the best-available
ages for each open cluster, we subtract the WD cooling age from the
cluster age to get the progenitor star age.  This is converted into
the progenitor's zero-age main sequence mass (the initial mass) using
Padova stellar evolutionary sequences \citep{Girardi2002} and observed cluster
metallicities \citep[For more detailed explanation of this
  process, see][]{Williams2006}.   This process leaves only the temperature
and surface gravity as potential inter-sample systematic error sources.

\subsection{Paring the Open Cluster White Dwarf Sample}
To further reduce uncertainties, we next apply some stringent
selection criteria to the WD sample.  First, we reject any WDs where
the uncertainty in either the open cluster age or the calculated
progenitor star lifetime is $\geq 50\%$.  Second, we reject any WDs
with masses less than $0.5M_\odot$, as these are likely products of
binary evolution, not the single-star evolution we wish to explore.
Lastly, we reject any WDs with final mass uncertainties greater than
10\% or cooling time uncertainties greater than 50\% (typically WDs
with low signal-to-noise observations).  Finally, we only include WDs
likely to be cluster members, with either proper motion determinations
or apparent distance moduli within 2$\sigma$ of the cluster distance
modulus.

The resulting sample includes 46 WDs from eight open clusters and two
binary-star systems: Sirius A/B \citep{Liebert2005} and Procyon A/B
(Liebert at al.~2006, in preparation).  We note that this is not
a complete sample of open cluster WDs, but as we
are not making any analysis of the cluster WD mass and/or luminosity
functions, this incompleteness should not make a significant difference.

\subsection{A Binned Initial-Final Mass Relationship}
For those clusters with multiple WDs, we now make the simplifying
assumption that all WDs in a given cluster have the same progenitor
mass and final mass.  We determine this cluster initial mass and
cluster final mass by finding the mean of the initial and final masses
of individual WDs in the cluster.  We define the scatter in each
quantity as the the standard deviation of the individual cluster WDs
about the cluster means.

The binning of a cluster's WD population into a single point has one
distinct advantage over previously-published IFMRs.  The largest
source of error in the initial mass is due to uncertainty in cluster
ages.  This error is systematic for WDs in a given cluster -- if the
cluster is older or younger, the WDs in that cluster will shift
together to lower or higher initial masses (respectively) in the IFMR.
For that reason it is not formally correct to add observational errors
in quadrature with the cluster age errors for each individual WD, as
done in the \citet{Ferrario2005}.  Yet errors due to cluster ages can
be considered random in cluster-to-cluster comparisons.  By binning
all points in the cluster, the errors due to cluster ages can be
represented as random errors in the IFMR.

We note that binning is not a formally correct tactic -- we expect
individual cluster WDs  to have a range of initial masses and final
masses.  However, for all clusters except M35 and Praesepe, we note
that the distribution of cluster WDs about the cluster means are
consistent with a Gaussian distribution.  Still, the assumption that
each cluster has a single initial and final mass needs to be relaxed
in further work on this topic.

\begin{figure}[!tb]
\begin{center}
\includegraphics[scale=0.5]{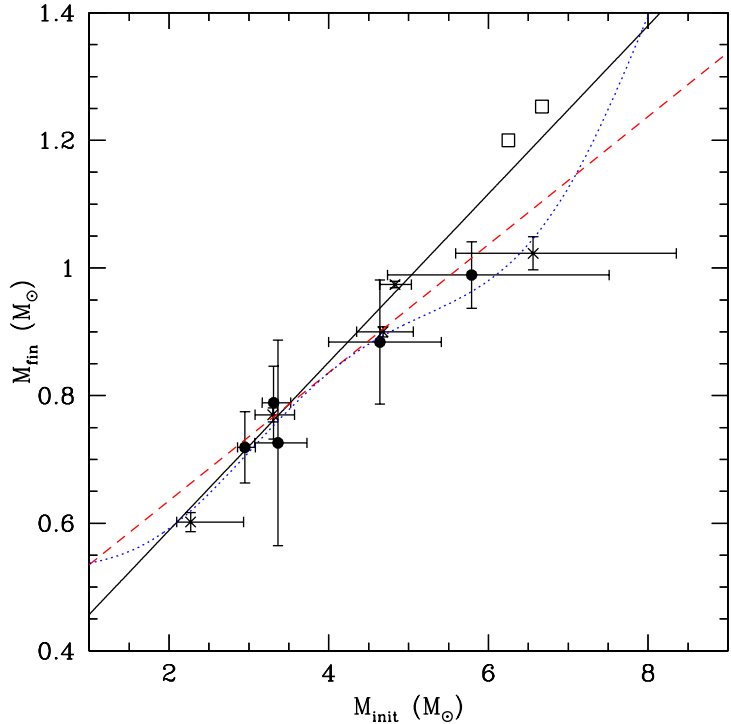}
\end{center}
\caption{The binned initial-final mass relation.  Filled circles are
  binned points from open clusters with four or more WDs; crosses are
  from clusters or binary systems with three or fewer WDs.  The solid
  line is a least-squares linear fit to these points. The dashed line
  is the linear fit from \citet{Ferrario2005}; the dotted line is the
  inversion of the field WD mass distribution presented in that work.
  Open squares, which were \emph{not} included in the fits, are from
  other work presented at this conference; namely, Dobbie et al.'s (2006b)
  points for GD~50 and PG~0136+251.  The
  agreement between these points and the extrapolation of the linear
  fit is encouraging. \label{fig.new_ifmr}}
\end{figure}

This binned initial-final mass relation is shown in Figure
\ref{fig.new_ifmr}. Comparison with Figure 1 shows that the
relationship appears much tighter.  Also shown in the Figure are the
linear relation from \citet{Ferrario2005} (dashed line) and their
relation obtained from inverting the field WD mass function (dotted
line).  Neither line is a poor fit, and the inflections at
high masses in the inverted fit mirror potential inflections in the
empirical relation.  We use least-squares fitting to obtain our own
linear fit to the binned IFMR:
$$M_{\rm f} =  (0.132 \pm 0.017) M_{\rm i}+0.33\pm 0.07\,.$$
This line is shown (solid) in Figure 2.

\section{Discussion}
Using this binned IFMR, we now explore some of the issues raised in
the introduction.

\emph{Is the IFMR linear or more complicated?} --- The $\chi^2$ value
of the linear fit to the binned IFMR, $\chi^2=2.74$
with 2 degrees of freedom, is an acceptable fit, though qualitatively
it appears that some curvature in the high-mass end of the IFMR may be
warranted.  Since the linear fit was calculated, new data have been
published for the high-mass WDs GD~50 and PG~0136+251, assuming both
are escaped members of the Pleiades \citep{Dobbie2006}.  These points
are shown in Figure 2 as open squares, and, despite the large error
bars (not shown), these points lie remarkably close to the linear
binned IFMR.  We also note that points presented by J.~Kalirai at this
conference lie close to extrapolation of this line at the low-mass end
of the relation.  We therefore conclude that there is no strong
evidence that the IFMR is non-linear at the current levels of
observational precision.

\emph{What is the intrinsic scatter in the IFMR?} --- We can estimate
the internal scatter in the IFMR based on four clusters with more than
5 WDs: M35, NGC 2099, Praesepe, and the Hyades.  We compare the stated
observational errors with the measured scatter for each cluster in the
binned IFMR (see Table 1).  In all four cases, the measured scatter is
larger than the stated errors, with an additional $0.05M_\odot$ of
scatter required.  The additional scatter may be due to
intrinsic scatter in the IFMR, but it could also be due to an
understatement of the observational errors or to systematic errors
caused by the assumption that all WDs in a given cluster have the same
initial mass and the same final mass.  As both of these will tend to
increase the intrinsic scatter, however, we can state with reasonable
confidence that the internal scatter of the IFMR is $\lesssim
0.05M_\odot$.

\begin{table}[!tb]
\caption{Stated Errors and Observed Scatter in Open Cluster Final Masses}
\smallskip
\begin{center}
{\small
\begin{tabular}{lccc}
\tableline
\noalign{\smallskip}
Cluster & $M_{\rm f}$ & Stated Error & Measured Scatter\\
& ($M_\odot$) & ($M_\odot$) & ($M_\odot$) \\
\noalign{\smallskip}
\tableline
\noalign{\smallskip}
Hyades   & 2.95 & 0.033 & 0.056 \\
Praesepe & 3.31 & 0.030 & 0.057 \\
NGC 2099 & 3.36 & 0.095 & 0.161 \\
M35 & 4.64 & 0.060 & 0.097 \\
\noalign{\smallskip}
\tableline
\end{tabular}
}
\end{center}
\end{table}

\emph{Is the IFMR metallicity-dependent?} --- Given that many
mass-loss mechanisms in evolved stars rely on metals and thus are
metallicity-dependent, it is reasonable to suspect that the IFMR is
metallicity dependent.  \citet{Kalirai2005}, using data from NGC 2099,
claim to find some indication that the IFMR may be dependent on
metallicity.  We briefly explore this by comparing the binned points
for three clusters of nearly-identical ages: the Hyades, with
$M_{\rm i}=2.95\pm 0.1,\, M_{\rm f}=0.72\pm 0.06$, and [Fe/H]$=0.14$
\citep{Perryman1998}; Praesepe, with $M_{\rm i}=3.31\pm 0.15,\, M_{\rm f}=0.79\pm
0.06$, and [Fe/H]$=0.14$ \citep{Claver2001}; and NGC 2099, with
$M_{\rm i}=3.4\pm 0.3,\, M_{\rm f}=0.73\pm 0.16$, and [Fe/H]$\approx -0.24$
\citep{Kalirai2005}.  Despite a difference in metallicity of $\approx
0.4 \, {\rm dex}$, there are no significant differences in $M_{\rm f}$
between these three points.  Therefore, any metallicity dependence of
the IFMR for $M_{\rm i}\approx 3 M_\odot$ must be smaller than $\Delta
M_{\rm f}\approx 0.05M_\odot$.

We again acknowledge that the quantitative conclusions above may be
affected quite markedly by our simplifying assumption that each
cluster can be represented by a single data point in the empirical
IFMR.  However, we feel that the data are still of sufficient quality
to make a few general conclusions.  Specifically, we conclude that the
IFMR appears to be linear within current observational limits over the
currently-observed range of initial masses.  We also conclude that
intrinsic scatter in the relation at a given initial mass is on the
order of or smaller than the observational errors and that there is no
compelling evidence for a metallicity dependence in the IFMR to within
the observational errors, at least for initial masses $\sim 3M_\odot$.
Further work on understanding the empirical IFMR and teasing out more
quantitative answers clearly remains.



\acknowledgements 

The following people all need to be thanked for reasons too numerous
to list here: Michael Bolte, James Liebert, Kate Rubin, Matt Wood,
Detlev Koester, Pierre Bergeron, Giles Fontaine, Lilia Ferrario, and
Dayal Wickramasinghe.  We are grateful for financial support from
National Science Foundation grant AST 03-07492 and an NSF Astronomy
and Astrophysics Postdoctoral Fellowship under award AST-0602288.


\end{document}